\documentclass[structabstract]{aa}  
\usepackage{graphicx,url}
\usepackage{txfonts}
\begin{document}
   \title{Detecting circumstellar disks around gravitational microlenses}

   \author{M. Hundertmark,
          F. V. Hessman \and S. Dreizler \fnmsep
   }

   \institute{Institute for Astrophysics, University of G\"ottingen,
              Friedrich-Hund-Platz 1, 37077 G\"ottingen\\
              \email{mhunder@astro.physik.uni-goettingen.de}
             }

\date{Accepted March, 2009}

\abstract
{}
{We investigate the chance of detecting proto-planetary or debris disks in stars that induce microlensing event (lenses).
The modification of the light curves shapes due to occultation and extinction by the disks as well as the additional gravitational deflection caused by the additional mass is considered.}
{The magnification of gravitational microlensing events is calculated using the ray shooting method. 
The occultation is taken into account by neglecting or weighting the images on the lens plane according to a transmission map of the corresponding disk for a point source point lens (PSPL) model. 
The estimated frequency of events is obtained by taking the possible
inclinations and optical depths of the disk into account.}
{We conclude that gravitational microlensing can be used, in principle, as a
  tool for detecting debris disks beyond 1\,kpc, but estimate that each year of the
  order of 1 debris disk is expected for lens stars of F, G, or K
  spectral type and of the order of 10 debris disks might have
  shown signatures in existing datasets.}
{}   
\maketitle
%

\section{Introduction}

Since the discovery of the first circumstellar dust disk around Vega by the \textit{Infrared Astronomy Satellite} (Aumann et al.\cite{aumann}), at least 101 circumstellar disks have been resolved around pre-main and main sequence stars out to a distance of 1 kpc\footnote{\url{http://www.circumstellardisks.org/}}, including 17 so-called debris disks.   
Proto-planetary disks may be quite common during the pre-main sequence life of a star, but with a maximum life span of only 10 Myr, a galactic star formation rate of $5\ \textrm{stars}/\textrm{yr}$, and an upper limit of 400 billion stars in the MilkyWay (McKee \& Williams \cite{mckee}, Diehl et al. \cite{diehl}), the chance of a random star containing a proto-planetary disk is only about 0.01 \%.
The debris disks that remain thereafter last much longer, perhaps as long as 10 Gyr (e.g. Greaves \cite{greaves}), and so should be much more common.
Trilling et al. (\cite{trilling}) have shown that the occurance of 70\,$\mu m$ excess emission in a sample of F, G, and K stars -- presumedly due to dust in some form of debris disk -- is $16\pm 3$\%.
In contrast, practically no main sequence M-stars show excess $70\,\mu m$ emission: Rhee et al. (\cite{rhee}) have found only one disk (AU Mic) from a sample of $\sim900$ M-stars.

While a common phenomenon, circumstellar disks around main sequence stars can
be difficult to detect using standard methods, coronography and the detection
of excess infrared emission.  A third method - detecting the disks in absorption against a
background source - works very well for dense proto-planetary disks seen
against an emission nebula ("proplyds"; McCaughrean \& O'Dell
\cite{mccaughrean}) but debris disks around main sequence stars are
optically thin and unlikely to be found in front of a bright background.

Gravitational lensing could provide an additional method for detecting and
characterizing circumstellar disks by creating a well-defined and bright if
temporary and geometrically complicated background source. If the projected distance between the lens star and the background source is small enough, the source star appears to separate into two geometrically extended objects -- the source of the photometric magnification in so-called microlensing, where the lensed object is not resolved.
The angular scale of gravitational lensing is given by the Einstein radius 
\begin{equation}
	\theta_E = \sqrt{ \frac{4GM}{c^2} \frac{D_{LS}}{D_S D_L} }.
	\label{eq_re}
\end{equation}
where $D_{LS}$ denotes the lens-source distance, $D_L$ the lens-observer distance, $D_S$ the source-observer distance and $M$ is the lens mass. 
For a typical galactic microlensing event observed towards the galactic bulge
(Paczy{\'{n}}ski \cite{paczynski}) with a $0.5\,M_{\odot}$ lens at $6\,kpc$
and a source at $8\,kpc$, one obtains an angular scale of the order of a $mas$
(Wambsganss \cite{wambsganss}), corresponding to a distance of $\sim2.5\,AU$
at the distance of the lens. Thus, the typical microlensing events studied primarily to detect low-mass exoplanets around lens stars (Beaulieu et al. \cite{beaulieu}; Bennett et al. \cite{bennett_b}) may also contain information about much more diffuse circumstellar matter as well.

 \begin{figure*}[htb]
   \centering
   \includegraphics{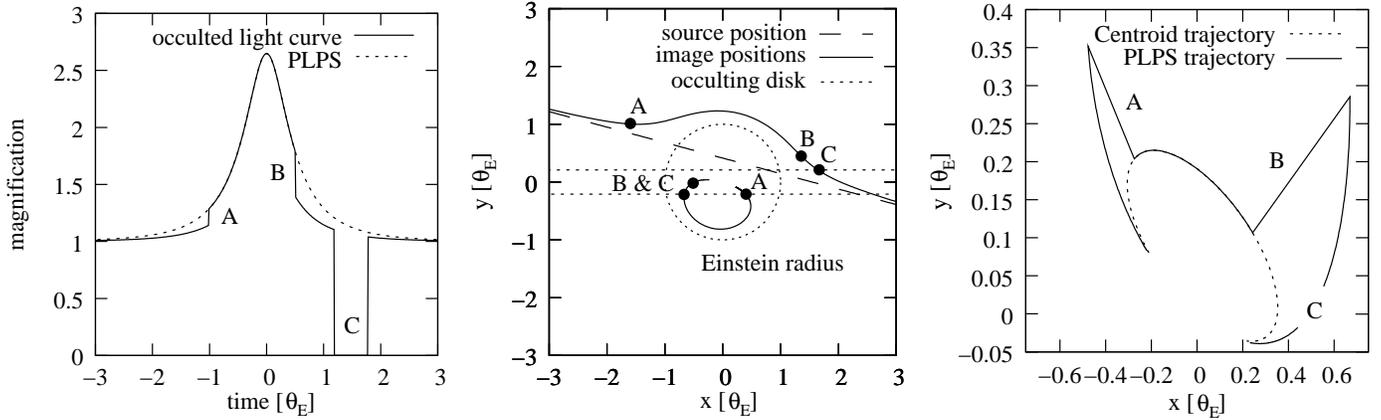}
   \caption{Light curve of the point source point lens model occulted by an
     optically thick edge-on disk is
     plotted along with the corresponding source and image tracks and the
     astrometrically detectable astrometric offset from the source. Features A and B are caused
     by the occultation of the less magnified image while feature C is obtained
     from the occultation of both images. The impact parameter is $0.4\ \theta_{\textrm{E}}$, the inclination $89.8\,^{\circ}$ and the disk is
     rotated by $15\,^{\circ}$.}
              \label{Fig_thick}%
    \end{figure*}

Bozza \& Mancini (\cite{bozza_a}) and Bozza et al. (\cite{bozza_b}) have
studied the effects of microlensing by lenses consisting of gas clouds and
interacting binaries which may be used as an approximation for a face-on proto-planetary disk and Zheng \& M{\'e}nard (\cite{zheng}) have considered
the effects of disks around microlensing source stars, but to our knowledge the detectibility
of circumstellar disks around lens stars has not been considered in the
literature. About $70\%$ of the observed lens stars
towards Baade's window are M stars and smaller and only $\sim25\%$ are of
spectral type F, G, or K (Dominik \cite{dominik_m}). 
Given that the microlensing survey groups OGLE (Udalski et al. \cite{udalski})
and MOA (Abe et al. \cite{abe}) are detecting about $1000$ events per year,
the fraction of events containing a debris disk should be significant: about
$4.0\%$ or currently $\sim40$/year.
The chance of detecting proto-planetary disks among lens stars is correspondingly lower: $0.0025\%$ or $\sim0.25$ per year (neglecting the additional absorption likely to be found within star-forming regions).
Although only 1 in 8000 lensed stars have proto-planetary disks,  future
surveys like the \textit{Microlensing Planet Finder} (Bennett et al.
\cite{bennett_a}) or \textit{EUCLID} (e.g. Beaulieu et al.\cite{beaulieu_b}) have a chance of detecting a significant number of such rare events.

The effects of a circumstellar disk on microlensing lightcurves can be separated into two very different regimes.  The disk can occult or at least extinct the light of the source star as it is bent around the lens star -- a simple geometric effect given a description of the geometry and matter distribution of the disk and the geodesic paths of the photons from the source.  
Alternatively, the mass of the disk itself can distort the geodesic paths of the photons and hence change the apparent areas of the distorted source images and the resulting lightcurve independent of any occultation or extinction effects.
The purely geometric disk parameters are the projected size, shape, inclination, and orientation angle $\phi$ of the major axis of the disk relative to the path of the source star and the size of the lens star's Einstein radius.
The typical outer radii of proto-planetary disks are about $100-800$\,AU and of debris disks about 50 to 100\,AU, (e.g. Greaves \cite{greaves})
corresponding to $20-320\,\theta_E$ for a typical lens. In both cases we
adapt an outer radius of 150\,AU as chosen by Brauer et al. (\cite{brauer}).

In the following we will discuss the chance of detecting circumstellar disks by simulating a variety of potential light curves. We will start with optically thick disks, where the optical depth is greater than one (cf. Krivov \cite{krivov}) representing the young proto-planetary disk phase and show that the corresponding magnification structure is scaled according to the ratio of the disk mass and its stellar host. This insight leads us to the most likely case of a microlens surrounded by an old debris disk modelled as optically thin and geometrically thick attenuating disks with optical depths $\tau \approx 0.01$.



\section{Optically thick proto-planetary disks}

Given that the major microlensing effects occur for impact parameters less than a few Einstein radii, a magnified background source will only be visible behind proto-planetary disks with high inclinations; disks with low inclinations may be too large to show any relativistic effects and will only show simpler occultations of non-magnified sources.
Similarly, debris disks with high inclinations are more likely to be detected because both the projected extinction and the projected mass-density responsible for any additional lensing effect will be higher.

An optically thick disk can be modeled as an occulting disk with given inclination and orientation angles as parameters. The features of the corresponding light curves and the astrometrically observable change of the centroid of both images are illustrated in Fig.~\ref{Fig_thick}, neglecting the gravitational deflection due to the disk mass. 
The less magnified image is occulted close to the maximum of
magnification. Because the disk structure is bound to the lens, it is unlikely
to occult the magnified image exclusively; unless the edge-on-disk is nearly parallel to the straight line through the two images and the lens. 
An asymmetric total occultation of both images as in Fig.~\ref{Fig_thick} indicates that an optically thick disk structure is present. 

In addition to the purely photometric effect, the occultation also
  produces a shift in the center-of-light position of the lensed source,
  albeit small. The astrometric offset due to the occultation of the less
  magnified images, without parallax effects (see Gould \cite{gould}), can be estimated by assuming that these images are located close to the Einstein radius
\begin{equation}
\delta \theta \approx \frac{\mu_{-}}{\mu} \theta_{\textrm{E}}=
\frac{1}{2} \left(1 - \frac{u \sqrt{u^2 + 4}}{u^2 + 2 }\right) 
\label{Eq_astrom_offset}
\end{equation}
where $u$ denotes the lens-source separation in $\theta_{\textrm{E}}$, $\mu$ the
total magnification and $\mu_{\pm}$ the magnification factor of magnified
and demagnified image. For a typical galactic microlensing event, the angular Einstein radius is
$\approx 0.4\,\textrm{mas}$: if the source-lens separation is below 1 Einstein
radius, the pre-factor in Eq.~\ref{Eq_astrom_offset} takes values between 0.12
and 0.5 and the total offset is of the order of 10\,$\mu\textrm{as}$ Fig.~\ref{Fig_thick}), which is
detectable using the planned \textit{GAIA} (Lindgren \& Perryman
\cite{lindgren}) or \textit{SIM} (Unwin et al. \cite{unwin})
satellites. For ground based observations Rattenbury \& Mao
  (\cite{rattenbury}) propose the usage of closure phase interferometry which
  could provide the desired accuracy in this context.

Given the rareness of such young disks and the difficulty of seeing
microlensing effects, detecting such disks photometrically and astrometrically
is not impossible but unlikely.


\section{Optically thin debris disks}

\subsection{Geometrically thin disks}

Up to now, we have neglected the gravitational deflection by the mass content of the circumstellar disk, represented by the mass surface density
\begin{equation}
\Sigma \approx 0.04\,g\,cm^{-2} ~ \left(\frac{q_D}{0.001}\right)
\left(\frac{M_\star}{0.5\,M_\odot}\right)
\left(\frac{R_D}{150\,AU}\right)^{-2} \left(\frac{r}{R_D}\right)^{-0.8}
\label{Eq_sigma}
\end{equation}
(e.g. Brauer et al. \cite{brauer}), where $q_D \equiv M_D / M_\star$  is the ratio of the disk mass to the stellar mass, and $R_D$ is the effective outer radius.

Assuming $q_D$ is small, the maximum change in deflection angle and magnification is for edge-on oriented disks, whose effective linear mass distribution can be calculated by integrating the surface density.
The deflection angle of the combined point source point lens (PSPL) and edge-on disk model has no rotational symmetry, so we have to consider the vectorial deflection angle $\vec{\alpha}$. 
An analytical expression can be derived if one integrates the normalized line
density in the lens plane $\lambda(\theta_{1})$ along the disk radius $\theta_D$, expressed in units of
the Einstein radius (cf. Narayan \& Bartelmann \cite{narayan}): 
\begin{equation}
\vec{\alpha}(\vec{\theta}) = \frac{1}{\left(1+q_D\right)}\cdot \frac{1}{\left|\vec{\theta}\right|} + 
\frac{q_D}{q_D+1} \int_{-\theta_D}^{\theta_D} \lambda(\theta_{1}')
\frac{\left(\vec{\theta}-\vec{\theta}'\right)}{\left|\vec{\theta}-\vec{\theta}'\right|^2}
\textrm{d}\theta_{1}'
\label{Eq_deflection_integral}
\end{equation}
The normalized line density for a mass distribution parallel to $\theta_1$ is given by
\begin{equation}
  \lambda(\theta_{1}) =  \int_{-\sqrt{\theta_D^2-\theta_{1}^2}}^{\sqrt{\theta_D^2-\theta_{1}^2}}
  \left(\theta_1^2+\theta_{3}'^2\right)^{-0.4} \textrm{d}\theta_{3}'
\end{equation}
where $\theta_{1,2}$ define the lens plane and the integral over $\theta_3$
projects the surface density distribution onto the line. The resulting expression contains the hypergeometric function $_2F_1$
(Abramowitz \& Stegun \cite{abramowitz}):
\begin{eqnarray}
 \lambda(\theta_{1}) & = & \left|\theta_1\right|^{1/5}
  \bigg[ 0.45 \bigg(22.18
  \left(\theta_D^2-\theta_1^2\right)^{1/10} \theta_1^{-2/10} \cdot
  \nonumber \\
&&  _2F_1 \left(-\frac{1}{10};\frac{4}{10};\frac{9}{10};\frac{\theta_1^2}{\theta_D^2-\theta_1^2}\right)-18.94 \bigg) \bigg]
\end{eqnarray}
\begin{figure}[!htb]
   \centering
   \includegraphics[width=0.47\textwidth]{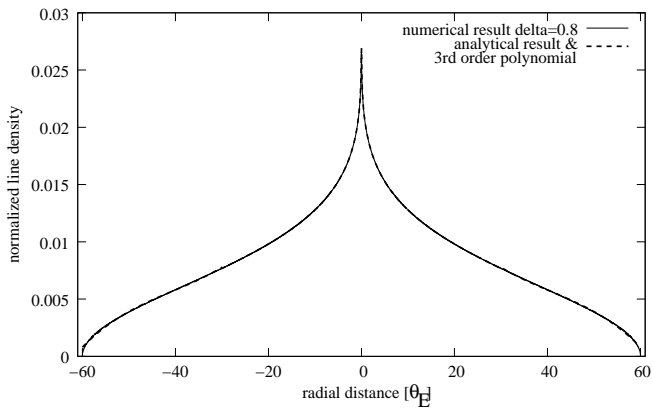}
   \caption{The normalized line density of an edge-on oriented disk with a surface density $\Sigma
   \propto \left(\frac{r}{1 \textrm{AU}}\right)^{-0.8}$, and a disk with $r_{\textrm{in}}=0.03\
\textrm{AU}$ and $\theta_D=60\,\theta_{\textrm{E}}$ is plotted.}
              \label{Fig_density_integration}%
\end{figure}
which can be efficiently evaluated sufficiently away from the singularities at $\pm
\theta_D$. When a third order polynomial fit is used
outside a 40\,$\theta_{\textrm{E}}$ radius, as shown in
Fig.\ref{Fig_density_integration}, the total deflection angle
(Eq.~\ref{Eq_deflection_integral}) can be numerically integrated. 

\begin{figure}[!htb]
   \centering
   \includegraphics[width=0.48\textwidth]{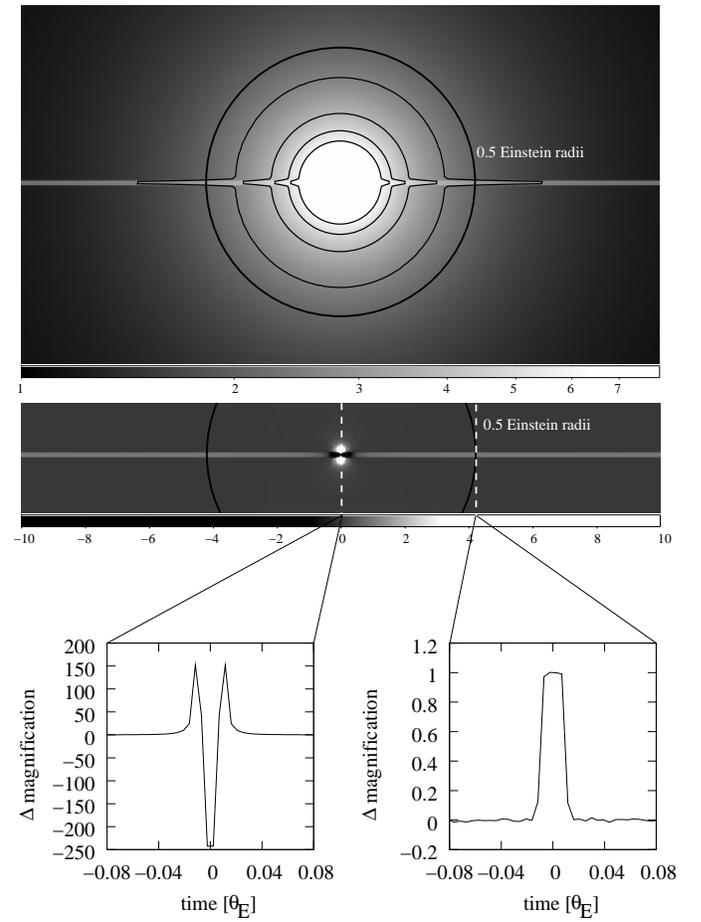}\\
   \caption{Top: magnification map for a linear mass distribution
   created by ray shooting (Eq.~\ref{Eq_deflection_integral}) for the combination of a point lens and a disk mass of 20 \% of the
   stellar host along with contour lines of magnifications at $\mu=2.5,\ 3.7,\ 4.8\
   \textrm{and}\ 6$. Profiles extracted from the magnification map located in
   the source plane correspond to light curves.  Middle: change relative to a
   point source point lens  (PSPL) model. Bottom: vertical cuts corresponding
   to deviations in the light curves for two perpendicular trajectories of the
   lens and source for $0, 0.5.\,\theta_{\textrm{E}}$}
              \label{Fig_magpattern}%
\end{figure}

Even though the lens equation with the additional term for the deflection angle cannot be solved analytically, the disk can be embedded into a ray shooting framework (Kayser et al. \cite{kayser}). For $q_D \sim0.2$, the calculated magnification offset within 1\,$\theta_{\textrm{E}}$ from the disk is
$\sim1$ and hence potentially observable. Close to the linear mass distribution representing the disk, one sees an additional rectangular magnification caused by
the attraction of rays (Fig.~\ref{Fig_magpattern}). For small source-lens separations $u$ (here $u<0.05\,\theta_E$) the singular peak splits in two at the central region. For $q_D \sim0.20$, the affected region of the
extended box is 0.02\,$\theta_E$ and the central distortion has a relative
deviation larger than 1\% for 0.2\,$\theta_E$, where the Einstein radius is calculated for the
total mass of the system.

According to Eq.~\ref{Eq_deflection_integral}, the
width of the distortion is proportional to $\lambda(\theta_{1})$ and thus to
the mass ratio $q_D$ for $q_D \ll 1$. 
The rectangular distortion is redistributed in case of an inclined disk
due to the conservation of rays: for $i=89^{\circ}$ the conserved integral of
the rectangular distortion reduces the peak deviation to 2\% and a face-on rotational symmetric disk has no effect at all. If the disk has
no central gap, a deviation containing a splitted singularity is observable as
long as the finite size of the source star does not smooth it out. Considering a bright giant source star with 10 $R_{\odot}$ and
thus $0.014\,\theta_E$ at 8\,kpc, this limit is reached at a mass ratio of
0.014 and, in case of inclined disks, even earlier. 

The calculation of the extinction in a geometrically and optically thin debris
disk requires an estimation of the vertical optical depth $\tau_\lambda (r) = \kappa_\lambda \Sigma(r)$. 
The mean R-band optical depth of AU\,Mic is in the range $10^{-3}$ to $10^{-4}$ (Kalas \cite{kalas}); using Eq.\,\ref{Eq_sigma} and an area-weighting of $\tau$, one obtains an estimate of the extinction coefficient (assumed to be constant)
\begin{equation}
\kappa_R \approx 0.016\,cm^2 g^{-1} ~ \left(\frac{\bar{\tau_R}}{10^{-3}}\right)  \left(\frac{q_D}{0.001}\right)^{-1}  \left(\frac{M_\star}{0.5\,M_\odot}\right)^{-1} 
\end{equation}
and a final R-band optical depth distribution
\begin{equation}
\tau_{R}(r) = 6\cdot10^{-4}  ~ \left(\frac{\bar{\tau_R}}{10^{-3}}\right) \left(\frac{R_D}{150\,AU}\right)^{-2} \left(\frac{r}{R_D}\right)^{-0.8}
\label{Eq_density}
\end{equation}
For a typical inner disk radius of $0.03\,AU$, $\tau_R$ is smaller than unity
even in the disk center. A thin disk with such low optical depths produces relative deviations of
the light curve of the order of 0.1\%. Current ground-based
microlensing observations require an order of magnitude higher deviation and
so could detect an average optical depth $>5 \cdot 10^{-3}$; according to
Augereau \& Beust (\cite{augereau}), this would be the maximal optical depth in the
visible for the AU Mic debris disk, transversed in vertical direction, but for
an edge-on orientation, optical depths of $4 \cdot 10^{-2}$ could be reached. 
Consequently edge-on configurations are much more likely to be detected from microlensing, too. 

\begin{figure*}[!htb]
   \centering
   \includegraphics[width=0.97\textwidth]{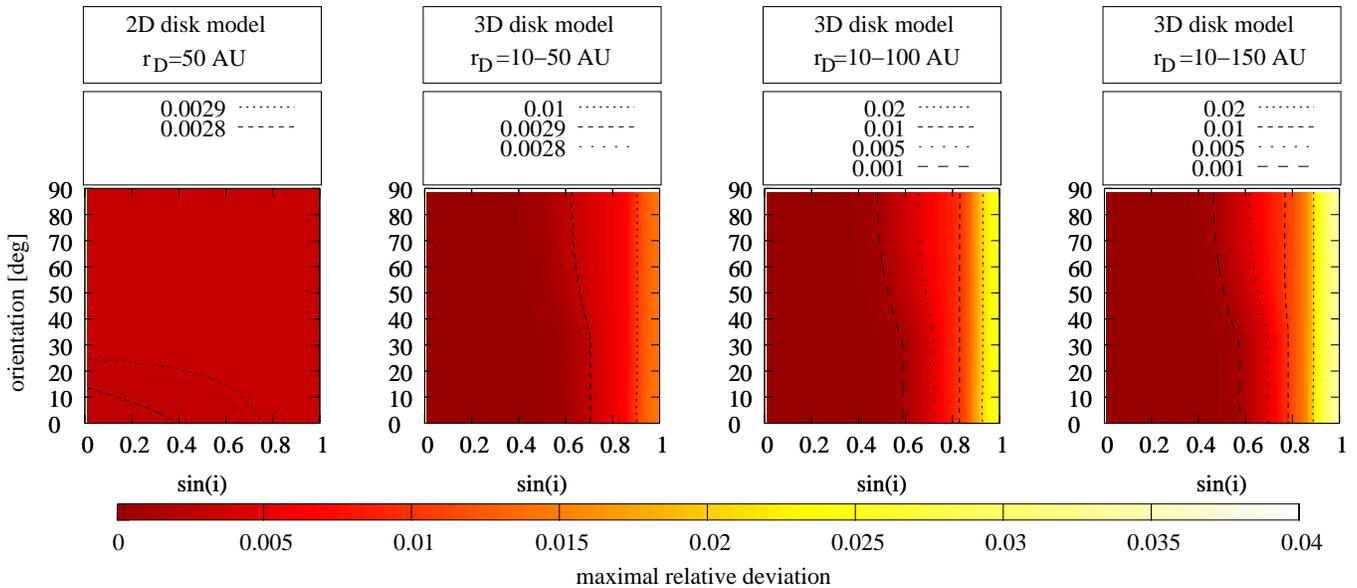}
   \caption{Contour lines of maximal relative deviation from PSPL and a grid of attenuated light
   curves with $u_0=0.1\,\theta_{\textrm{E}}$ are plotted for a surface
   density $\Sigma \propto r^{-0.8}$, different outer disk
   radii $r_D=50,\ 100,\ 150\,AU$, an inner disk radius of 10\,AU, and a constant mean optical depth of
   $\bar{\tau} =5 \cdot 10^{-4}$ }
   \label{Fig_smap}
\end{figure*}

\subsection{Geometrically thick disks}

The attenuating effect of a circumstellar disk depends especially on the path
length of the light ray through the disk and thus on its shape. For an edge-on
configuration and a 0.5\,$M_{\odot}$ star of 0.6\,$R_{\odot}$, the deflection causes a
vertical shift of $10^{-3}\,AU$ at 150\,AU in the vertical direction and can be
neglected. In the following, a geometrically thick disk with a gaussian vertical dust distribution and a scale height $H(r)=0.47\left(\frac{r}{50\,AU}\right)+0.34\,AU$ is used (for AU Mic cf. Metchev et al. \cite{metchev} and
Krist et al. \cite{krist}). The density $\rho$ of the disk is given by
\begin{equation}
 \rho(r,z)= \frac{\Sigma(r)}{H(r) \sqrt{2\pi}} e^{-z^2/2H(r)}
\end{equation}
%
%
%
The normalization constant is determined by integrating $\rho$ parallel to the
disk normal. 
%
\begin{equation}
\tau_{\textrm{R}}=5.07\cdot 10^{-6} \int_{0}^{s} \rho(r,z) \textrm{d} s
\end{equation}
In PSPL models the light path is usually parameterized using two straight lines according to the
thin lens approximation: To first order, the true light path is a
hyperbola: according to G\"onner (\cite{goenner}, p. 206), the maximal deviation of the distance of closest approach $R$ from the hyperbola is
\begin{equation}
\Delta R = \frac{2 G M R}{R c^2 + 2 G M}
\end{equation}
where $M$ is the mass of the deflecting star. The maximal deviation from the thin lens approximation for
grazing incidence is $10^{-5}\,\textrm{AU}$ at $0.6\,R_{\odot}$ and a deflector mass of $0.5\,M_{\odot}$.

The sensitivity of different models for $\tau \sim5 \cdot 10^{-4}$ is shown in
Fig.~\ref{Fig_smap}. The source track for each pair of $(i,\phi)$ was limited to a range from
-1 to 1 $\theta_{\textrm{E}}$ and the impact parameter
$u_0=0.1\,\theta_{\textrm{E}}$ has been kept fixed; it does not change the maximal variation of the transmissivity, only
the maximal distance from the center. While the two dimensional model (projected surface density) with $\Sigma \propto
r^{-0.8}$ shows a maximal deviation below 1\%, the more realistic
three dimensional model with $\Sigma \propto r^{-0.8}$ shows an increased
deviation at high inclinations, because the transversed matter is taken into
account. The smaller the outer radius of the disk is, the poorer the
detectability. From the calculated sensitivity pattern the fraction of the
parameter space exceeding 1\%, i.e. 10\,mmag, can be calculated: 17\% for
$r_D=50\,AU$, 34\% for $r_D=100\,AU$  and 50\%
for $r_D=150\,AU$. These results are valid for a system without an inner
clearing but Wyatt et al. (\cite{wyatt}) have concluded from observations that
$2\pm2\%$ show hot dust close to the star. Assuming an inner radius of
10\,AU changes the detectability to 10\% for $r_D=50\,AU$ , to 18\% for
$r_D=100\,AU$ and to 23\% for $r_D=150\,AU$. Changing the scale height redistributes the sensitivity to smaller inclinations.

\begin{figure*}[!htb]
   \centering
   \includegraphics[width=0.95\textwidth]{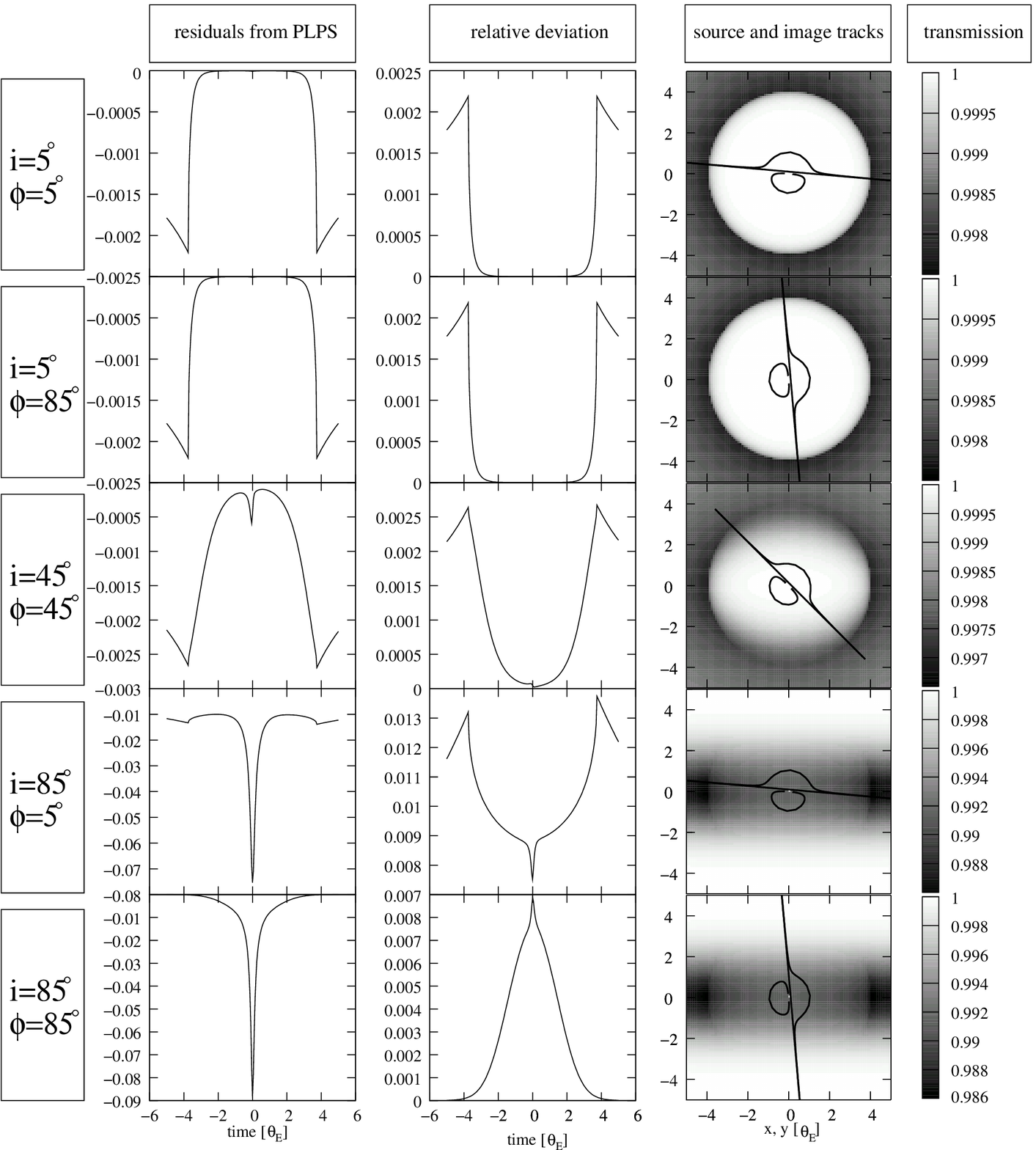}
   \caption{Differential light curves of a geometrically thick disk with a central hole are shown for different inclinations $i$ and orientations
   $\phi$ of the source track. The disk has an inner radius of 10\,AU and an
   outer radius of 50\,AU,  $\Sigma \propto
   r^{-0.8}$; $H=2.5\,\textrm{AU}$, and the average optical depth $\tau \approx
   5\cdot10^{-4}$}
              \label{Fig_3d_light_curves}%
    \end{figure*}

In Fig.~\ref{Fig_3d_light_curves} five different sensitivity configurations are calculated for $r_D = 50\,AU$ and an inner gap with
$10\,AU$ radius. Simulations for 100\,AU and 150\,AU show similar results: all of them show a systematic decrease in the residuals close to the maximal magnification, especially at high inclination. The source and image tracks plotted over the map of
transmissivity explain this behavior for a surface density modeled as
an inverse power law. The image with positive parity reaches its maximal separation from
the lens when the source-lens distance is minimal and therefore the optical
depth $\propto \Sigma$ is also minimal. This indicates one possibility of detecting circumstellar disks - fitting a PSPL model
and analyze the structure of the residuals within a few Einstein radii. The
simulated light curves show deviations between 0.1\%
and 1\% (i.e. 1 and 10\,mmag without blending flux), a signal that is
detectable by a dedicated space mission like the \textit{Microlensing Planet
  Finder} or \textit{EUCLID}. 

\section{Relative extinction curves}
It is possible to tell a priori if a certain microlens is surrounded by
attenuating matter if a wavelength-dependent extinction signal can be seen. When fitting a
PSPL model, the residuals can be used to show where relative attenuation in the
lens plane occurs between images taken at different wavelengths. A combination of photometric and astrometric follow-up
could then infer the extinction structure of the central zone by measuring
the extinction coefficient for each image: the centroid of the lensed source star will move
slightly to the less dimmed image. The centroid $\vec{\theta}_c$ with
extinction is located at
\begin{equation}
\vec{\theta}_c=\frac{k_{+} \vec{\theta}_{+} \mu_{+} + k_{-} \vec{\theta}_{-} \mu_{-} }{k_{+} \mu_{+} + k_{-} \mu_{-}}
\label{Eq_centroid}
\end{equation}
where $k_{\pm}$ denote the measured transmission coefficients, $\mu_{\pm}$ the
unextincted magnifications, $\vec{\theta{\pm}}$ the image positions and $\vec{\theta}_c$ the trajectories on the sky. The
total magnification
is described as
\begin{equation}
\mu = k_{+} \mu_{+}+k_{-}\mu_{-}
\label{Eq_magtotal}
\end{equation}
One of the transmission coefficients can be calculated from Eqs. \ref{Eq_centroid} and \ref{Eq_magtotal}:
\begin{equation}
k_{-}= \frac{\mu}{\mu_{-}} \frac{\vec{\theta}_c -\vec{\theta}_{+}}{\vec{\theta}_{-}-\vec{\theta}_{+}}
\label{Eq_extinct}
\end{equation}
Ideally brightness and position are measured
simultaneously and thus the measured magnification $\mu$ and position $\vec{\theta_c}$ can be jointly used in Eq.~\ref{Eq_extinct}, after correcting
for blended flux and baseline. Finally $k_{+}$ is calculated
from Eq.~\ref{Eq_magtotal} by using the image positions $\vec{\theta}_{\pm}$ from
the fitted PSPL model.

%
%
%
%
 
\section{Conclusions}

We have shown that gravitational microlensing can - in principal - be used as a tool for
detecting circumstellar disks beyond 1\,kpc when the photometric
residuals relative to a PSPL model are searched for deviations between -3 and 3\,$\theta_{\textrm{E}}$ around the maximal magnification. We estimate that 4\% of all
F, G \& K-microlenses contain debris disks. However, if the optical depth $\tau \geq
5\cdot 10^{-4}$ (age $\sim$ 1\,Gyr), a tenth of these objects should have
detectable disks. Assuming a power law model for the
surface density $\Sigma \propto r^{-0.8}$ and depending on the inner and outer
radii distribution of debris disks between 10 and 50\% of all remaining light curves
could show deviations larger than 1\%. For $\sim4000$ existing light curves from the
OGLE collaboration, we expect that of the order of 10 of the F, G \& K-lenses show a detectable
signature induced by a debris disk. If we take into account that anomalous
microlensing events, i.e. events that are not well-described by a PSPL model,
can also display systematic variations and blending occurs, it is unlikely, that debris disks can
be discovered from the existing datasets. Nevertheless, future space missions will almost
certainly provide the required accuracy for detecting debris disks and probing
the attenuating matter.


\begin{acknowledgements}

This work was inspired by the Wilhelm und Else Heraeus Physics School ``The Early Phase of Planet Formation''. 
M.H. would like to acknowledge the support by the German-Israelian Foundation and from the Graduiertenkolleg 1351.

\end{acknowledgements}

\end{document}